\documentclass[%
 aip,
 amsmath,amssymb,
 reprint,%
]{revtex4-1}

\usepackage{graphicx}
\usepackage{dcolumn}
\usepackage{bm}
\usepackage{dsfont} 
\usepackage[hidelinks]{hyperref} 
\hypersetup{
    colorlinks=true,
    linkcolor=blue,
    filecolor=blue,      
    urlcolor=blue,
    citecolor=blue,
}

\begin{document}

\preprint{AIP/123-QED}

\title{Search for long lasting electronic coherence using on-the-fly ab initio semiclassical dynamics}

\author{Alan Scheidegger}
\affiliation{Laboratory of Theoretical Physical Chemistry, Institut des Sciences et Ing\'enierie Chimiques, Ecole Polytechnique F\'ed\'erale de Lausanne (EPFL), CH-1015 Lausanne, Switzerland}

\author{Ji\v{r}\'{\i} Van\'{\i}\v{c}ek}
\affiliation{Laboratory of Theoretical Physical Chemistry, Institut des Sciences et Ing\'enierie Chimiques, Ecole Polytechnique F\'ed\'erale de Lausanne (EPFL), CH-1015 Lausanne, Switzerland}

\author{Nikolay V. Golubev}
\email{nik.v.golubev@gmail.com}
\affiliation{Laboratory of Theoretical Physical Chemistry, Institut des Sciences et Ing\'enierie Chimiques, Ecole Polytechnique F\'ed\'erale de Lausanne (EPFL), CH-1015 Lausanne, Switzerland}

\date{\today}

\begin{abstract}
Using a combination of high-level \textit{ab initio} electronic structure methods with efficient on-the-fly semiclassical evaluation of nuclear dynamics, we performed a massive scan of small polyatomic molecules searching for a long lasting oscillatory dynamics of the electron density triggered by the outer-valence ionization. We observed that in most of the studied molecules, the sudden removal of an electron from the system either does not lead to the appearance of the electronic coherence, or the created coherences become damped by the nuclear rearrangement on a time scale of a few femtoseconds. However, we report several so far unexplored molecules with the electronic coherences lasting up to 10~fs which can be good candidates for experimental studies. In addition, we present the full-dimensional simulations of the electronic coherences coupled to nuclear motion in several molecules which were studied previously only in the fixed nuclei approximation.
\end{abstract}

\pacs{Valid PACS appear here}
\maketitle

\section{Introduction}
The recent progress in laser technologies~\cite{Krausz_Ivanov:2009,Calegari_Nisoli:2016} made it possible to study properties of matter with unprecedented resolution. Ultrashort intense laser pulses revolutionized the field of atomic and molecular physics~\cite{Nisoli_Martin:2017} providing the scientific community with a unique tool to initiate and trace dynamics of both electrons and nuclei of a molecule in real time and with atomic spatial resolution.~\cite{Leone_Vrakking:2014,Nisoli:2019} 

Exposing a molecule to a short light pulse may bring the system to a nonstationary quantum state, thus launching a coupled dynamics of electronic and nuclear wave packets. Coherent superposition of multiple electronic states often results in ultrafast evolution of molecular observables, such as the electron density. Although driven by purely electronic effects,~\cite{Cederbaum_Zobeley:1999,Kuleff_Cederbaum:2014} the dynamics of the electron density is strongly coupled to the nuclear motion in the system. It was demonstrated in numerous studies~\cite{Vacher_Robb:2015,Vacher_Robb:2015a,Jenkins_Robb:2016,Jenkins_Robb:2016a,Arnold_Santra:2017,Vacher_Malhado:2017,Despre_Kuleff:2018} that the slow nuclear rearrangement has a dramatic impact on the electronic dynamics which leads to the fast decoherence of electronic oscillations on a time scale of just a few femtoseconds.

Capturing the interplay between nuclear rearrangement and the ultrafast electron motion requires a concerted description of the electron-nuclear dynamics. One of the most powerful approaches allowing the accurate description of this truly \textit{molecular} quantum dynamics is the multi-configurational time-dependent Hartree (MCTDH) method.~\cite{Meyer_Cederbaum:1990,Beck_Meyer:2000} MCTDH has been used successfully for simulating nonadiabatic wave packet dynamics in molecules~\cite{Raab_Cederbaum:1999} and recently applied for the description of electronic coherence.~\cite{Arnold_Santra:2017,Despre_Kuleff:2018} Although this rigorous technique, on the one hand, makes it possible to take into account all the quantum effects, such as tunneling and nonadiabatic transitions, on the other hand, it suffers from an exponential scaling problem and also requires the costly construction of global potential energy surfaces (PESs).

To overcome the constraints of techniques utilizing the precalculated form of PESs, methods evaluating the electronic structure ``on the fly'' have been developed. These ``direct dynamics'' approaches calculate the PESs only along trajectories, thus sampling only the relevant regions of the configuration space and avoiding the precomputation of globally fitted surfaces. Direct dynamics techniques range from fully quantum methods, such as the variational multi-configurational Gaussians (vMCG),~\cite{Burghardt_Meyer:1999,Worth_Burghardt:2003} \textit{ab initio} multiple spawning,~\cite{Curchod_Martinez:2018} coupled coherent states,~\cite{Shalashilin_Child:2004} multi-configurational Ehrenfest,~\cite{Saita_Shalashilin:2012} and Gaussian dephasing representation,~\cite{Sulc_Vanicek:2013} to more approximate mixed quantum-classical and semiclassical approaches including, e.g., the surface hopping,~\cite{Tully:1990} Ehrenfest dynamics,~\cite{Li_Frisch:2005} and Herman--Kluk propagator,~\cite{Herman_Kluk:1984} together with its extensions.~\cite{Tatchen_Pollak:2009,Ceotto_Atahan:2009a} Both the approximate Ehrenfest-based schemes~\cite{Vacher_Robb:2015,Vacher_Robb:2015a,Jenkins_Robb:2016,Jenkins_Robb:2016a} and the numerically exact vMCG method~\cite{Vacher_Malhado:2017,Jenkins_Robb:2018} have been used recently for computing the influence of nuclear structure and dynamics on the electronic coherences.

Despite the success of both MCTDH and trajectory-guided direct dynamics techniques in accurate description of the electronic dynamics coupled to nuclear motion, the above mentioned methods are still rather expensive computationally and thus only a few relatively small molecules have been studied so far. The very limited number of molecules analyzed to date is insufficient to make clear conclusions on how the molecular structure and the presence of specific functional groups influence the time scale of electronic coherence. At the same time, the possibility to measure experimentally the ultrafast electron motion in a molecule depends crucially on the number of oscillations which the electron density has time to perform before the electronic coherence is destroyed by the nuclear rearrangement.~\cite{Belshaw_Greenwood:2012,Calegari_Nisoli:2014} Therefore, computational preselection of molecules having the desired properties, including the long lasting electronic coherence, is essential for the successful experimental studies.~\cite{Lepine_Vrakking:2014}

Interestingly, the high-level quantum and multi-trajectory methods are somewhat redundant for the treatment of electronic coherence in polyatomic molecules. In most of the previously studied systems, the electronic coherence becomes suppressed by the slow nuclear rearrangement within the first ten femtoseconds of dynamics. This very short time scale typically implies that the nuclei remain very close to their original positions. Accordingly, the simulations of ultrafast electronic dynamics under the influence of nuclear motion require the capturing of only the first initial instant of molecular rearrangement. The latter makes the usage of expensive high-level techniques, initially designed for simulating long lasting nuclear dynamics, both unnecessary and impractical for systems, in which one is interested primarily in the evolution of the electronic subsystem.

Recently,~\cite{Golubev_Vanicek:2020} it was demonstrated that a simple semiclassical approach,~\cite{Heller:1975} in which the propagating nuclear wave packet is approximated by a single Gaussian function, can compete in accuracy with the full-dimensional quantum techniques for calculating the electronic coherence. In this semiclassical approach, the center of the Gaussian follows classical Hamilton's equations of motion while the width and the phase of the wave packet are propagated using the local harmonic approximation of the PES. Because the width of the Gaussian evolves in time, the approach was termed the thawed Gaussian approximation (TGA) in the literature.~\cite{Heller:1975,Begusic_Vanicek:2018a,Lasser_Lubich:2020} Importantly, the TGA gives the exact solution of the time-dependent Schr\"odinger equation in the limit when the propagation time approaches zero and thus is particularly suited for the treatment of processes taking place on short time scales.

Here, we use on-the-fly \textit{ab initio} semiclassical TGA to perform a massive scan of small polyatomic molecules searching for the long lasting electronic coherence. We concentrate on studying the electronic dynamics triggered by ionization of a molecule because the real-time tracing of a positive charge is currently more affordable from the experimental point of view.~\cite{Belshaw_Greenwood:2012,Calegari_Nisoli:2014,Kraus_Woerner:2015,Laraastiaso_Martin:2018} Furthermore, we preselect only those molecules which demonstrate special features in their ionic spectra: namely (I) the presence of a large energy gap between the low-energy valence ionic states and the remaining ones, which can be important for the experimental possibility to create the desired superposition of only a limited number of electronic states, and (II) the presence of a strong correlation between valence electrons of the neutral molecule, the so-called hole-mixing phenomenon,~\cite{Breidbach_Cederbaum:2003} which guarantees the existence of non-trivial dynamics of the electron density resulting from the created superposition.

We use a free online database PubChem~\cite{Kim_Bolton:2019} maintained by the National Institute of Health as a source of molecules in our search procedure. PubChem contains millions of molecular structures with additional information, such as physical properties, toxicity data, molecular identifiers etc. The database has an efficient search interface to easily access molecules with the desired properties and allows one to download the chemical structures in many standard formats. We limit the search procedure to consider the subset of small neutral molecules composed of C, H, O, and N atoms. Since thousands of molecules fulfill the aforementioned conditions, we randomly select about 250 molecules for the subsequent simulations. Therefore, the major goal of this study is not to test all the molecules in the database but to propose an efficient procedure to scan a large variety of molecules and to find systems best suited for the experimental investigations.

In addition to searching for long lasting electronic dynamics in so far not investigated molecules, we perform calculations of electronic coherence in several experimentally interesting systems which were previously studied employing the fixed nuclei approximation, namely 2-propyn-1-ol,~\cite{Breidbach_Cederbaum:2007} 2-phenylethyl-N,N-dimethylamine (PENNA),~\cite{Lunnemann_Cederbaum:2008} 3-buten-N,N-dimethylamine (BUNNA),~\cite{Lunnemann_Cederbaum:2008a} and 3-methylen-4-penten-N,N-dimethylamine (MePeNNA).~\cite{Lunnemann_Cederbaum:2008a} We show that the electronic coherence in these molecules becomes suppressed by the nuclear rearrangement within first few femtoseconds of the dynamics which is still enough to observe at least one clear oscillation of the electron density along a molecular chain. 


The remainder of the paper is organized as follows. In Sec.~\ref{sec:theory} the theoretical basis of the methods used to compute electronic coherences is presented. In Sec.~\ref{sec:comp_det}, we discuss the methodology and computational details of the approaches used in this study. The results of our simulations, including the discovered molecules with interesting properties, are presented in Sec.~\ref{sec:results}. In Sec.~\ref{sec:conclusions}, we summarize the results and conclude.

\section{Theoretical background}
\label{sec:theory}
The starting point of our simulations is a closed-shell neutral molecule in its ground electronic and vibrational state. Applying an intense laser pulse with an appropriate energy range, one can kick out an electron from the molecule and produce an ion. Explicit modeling of the ionization process requires simultaneous accounting for both the long-range continuum wavefunction of the leaving electron and the intricate correlated short-range structure of the bound electrons which is currently beyond the reach for all but the smallest systems.~\cite{Marante_Martin:2014,Marante_Martin:2017,Ruberti:2019} A practical way to describe the ionization process in realistic molecules is to employ the sudden approximation,~\cite{Cederbaum_Domcke:1977} i.e., assuming an instant removal of an electron from the system. In this case, the ionic state is prepared by projecting the electronic state of the neutral molecule onto the ionic subspace of the system. Furthermore, we employ the Franck--Condon approximation assuming that the nuclear wavefunction of the neutral ground state remains unchanged during its transfer to the ionic subspace.

After ionization, a single nuclear Gaussian wave packet on each involved ionic surface is propagated independently from the others. In our simulations, we concentrate only on those molecules in which the energy gaps between the ionic states remain large enough to justify the negligibility of the nonadiabatic population transfer.~\cite{Domcke_Koppel:2004} This condition is controlled by tracing the energies of the corresponding ionic states along each of the propagated trajectories. Although this scheme inevitably leads to the omission of potentially interesting molecules, it allows us to guarantee the applicability of the TGA and thus high accuracy of the reported results.

In Sec.~\ref{sec:elec_structure}, we review briefly the electronic structure methods which we use for computing the ionic states. In Sec.~\ref{sec:sudden_ioniz}, we describe the sudden ionization approximation and the hole-mixing mechanism. Section~\ref{sec:charge_migr} discusses a general procedure for computing observables. In Sec.~\ref{sec:TGA}, we present the equations of motion for the evolution of a Gaussian wave packet within the TGA. The theory part ends with Sec.~\ref{sec:SC_analysis}, where we describe in detail a simple phase-space approach allowing us to analyze the electronic coherence in physically intuitive terms.

\subsection{Electronic structure approaches}
\label{sec:elec_structure}
The centerpiece of our simulations is an accurate description of electronically excited states. As was explained in the introduction, we concentrate on studying the electronic dynamics in singly ionized molecules. From the computational point of view, the key difference in the description of neutral and ionized systems is the number of unpaired electrons in the ground electronic state. While most of the neutral molecules have an even number of electrons, thus forming a closed-shell ground electronic configuration, the removal of an electron during the ionization leads to the appearance of an open-shell species. Accurate description of the electronically excited states of an open-shell molecule is often case specific~\cite{Krylov:2017} and thus is not well suited for the automatic search procedure.

The latter difficulty can be efficiently circumvented by utilizing electronic structure approaches designed to describe $(N-1)$-electron system starting from the $N$-electron reference ground state. To this end, several computational schemes, such as the outer-valence Green's function (OVGF) method,~\cite{Cederbaum_Domcke:1977,Niessen_Cederbaum:1984} two-particle-one-hole Tamm-Dancoff approximation (2ph-TDA),~\cite{Niessen_Cederbaum:1984} a so-called partial third-order (P3)~\cite{Ortiz:1996} and its renormalized variants~\cite{Ortiz:2005} schemes, as well as the algebraic diagrammatic construction (ADC)~\cite{Schirmer:1982,Schirmer_Walter:1983} and the equation-of-motion coupled-cluster for ionization potential (EOM-CC-IP)~\cite{Stanton_Bartlett:1993,Gour_Piecuch:2006,Kamiya_Hirata:2006} techniques, have been developed in last few decades. Among these methods, ADC and EOM-CC-IP have been proven to be both highly accurate and computationally efficient~\cite{Schneider_Trofimov:2015,Dempwolff_Dreuw:2019,Ranasinghe_Bartlett:2019} which make them appropriate techniques for studying electronically excited states of ionized molecules.

The main weakness of the ADC and EOM-CC-IP techniques is the computation of nuclear gradients and Hessians required for the propagation of the thawed Gaussian wave packets. While nuclear gradients are available in EOM-CC-IP method at a large computational cost,~\cite{Stanton_Gauss:1994} the procedure for computing analytically nuclear gradients in ADC and Hessians in both ADC and EOM-CC-IP has not been reported yet. For that reason, we utilize the time-dependent density-functional theory (TDDFT)~\cite{Runge_Gross:1984,Casida:1995} for a fast initial pre-screening of the electronic coherences created by ionization. The TDDFT is used only for a preliminary scan of large number of molecules and finding promising candidates, which are then re-analyzed using high-level \textit{ab initio} ADC and EOM-CC-IP methods.

\subsection{Sudden ionization and hole-mixing mechanism}
\label{sec:sudden_ioniz}
Within the sudden approximation, the ionization event, i.e., the removal of an electron from a molecule, takes place on an infinitely short time scale. In practice, the time needed for the remaining electrons to respond to a sudden ionization was found to be about 50 attoseconds.~\cite{Breidbach_Cederbaum:2005} It was shown~\cite{Breidbach_Cederbaum:2005} that this time is universal, i.e., it does not depend on the particular system, and as such appears as the time scale of the electron correlation. Therefore, experimentally the sudden ionization can be achieved by applying an intense high-energy pulse with a duration shorter than the electron correlation time.

Mathematically, the removal of an electron from a system can by modeled by applying the annihilation operator $\hat{c}_{i}$ to the ground neutral state
\begin{equation}
\label{eq:psi0_sudden}
	|\Psi^{N-1}_{i}(0)\rangle=\hat{c}_i |\Psi^{N}_0 \rangle,
\end{equation}
where index $i$ refers to a particular molecular orbital of the neutral system and $|\Psi^{N-1}_{i}(0)\rangle$ is the initial state of the obtained ion at time $t=0$. The initial state $|\Psi^{N-1}_{i}(0)\rangle$ is, in general, not an eigenstate of the cationic Hamiltonian and thus will evolve in time. The ultrafast multielectron dynamics triggered by the ionization can manifest as the time-evolution of the created hole along a molecular chain. Driven by purely electronic effects, this mechanism was termed ``charge migration''~\cite{Cederbaum_Zobeley:1999,Kuleff_Cederbaum:2014} to distinguish it from a more common charge transfer driven by nuclei.~\cite{May_Kuhn:2003,Sun_Remacle:2017}

To understand the subsequent dynamics, let us expand the initial electronic wavefunction in a basis of ionic eigenstates $|\Psi_I^{N-1}\rangle$:
\begin{equation}
\label{eq:psi0_expanded}
	|\Psi^{N-1}_{i}(0)\rangle = \sum_I a_{I} |\Psi_{I}^{N-1}\rangle,
\end{equation}
where $a_{I}$ are the expansion coefficients. Expanding the cationic eigenstates in a configurational interaction series,
\begin{equation}
\label{eq:app_cat_expansion}
	|\Psi_I^{N-1}\rangle=\sum_k c_{k}^{I} \hat{c}_k |\Phi_0^N\rangle
	+\sum_{k<l} \sum_{a} c_{kla}^{I} \hat{c}^{\dagger}_a \hat{c}_k \hat{c}_l |\Phi_0^N\rangle + \cdots,
\end{equation}
where $c_{k}^{I}$ and $c_{kla}^{I}$ amplitudes are known from \textit{ab initio} calculations, one can solve the system of linear equations for the unknown coefficients $a_{I}$ in Eq.~(\ref{eq:psi0_expanded}). To illustrate the procedure, we consider an idealized model situation when only two strongly correlated electrons occupying orbitals $i$ and $j$ of a neutral system contribute to the resulting ionic states. In this case, the electron correlation leads to the appearance of hole-mixing between corresponding orbitals in the ionic states~\cite{Breidbach_Cederbaum:2003}
\begin{equation}
\begin{split}
	|\Psi_I^{N-1}\rangle=&c_1 \hat{c}_i |\Psi_0^N\rangle + c_2 \hat{c}_j |\Psi_0^N\rangle,\\
	|\Psi_J^{N-1}\rangle=&c_2 \hat{c}_i |\Psi_0^N\rangle - c_1 \hat{c}_j |\Psi_0^N\rangle,
\end{split}
\end{equation}
where, due to the orthonormality of the states, the two real coefficients $c_1$ and $c_2$, representing hole-mixing amplitudes, satisfy the relation $c_1^2 + c_2^2=1$. The electronic wave packet formed from these two states at time $t=0$ reads
\begin{equation}
	|\Psi^{N-1}_{i}(0)\rangle = a_{I} |\Psi_I^{N-1}\rangle + a_{J} |\Psi_J^{N-1}\rangle,
\end{equation}
which can be expressed in terms of the ionization out of the ground state as
\begin{equation}
\label{eq:psi0_states}
\begin{split}
	|\Psi^{N-1}_{i}(0)\rangle 
	= &(a_{I} c_1 + a_{J} c_2) \hat{c}_i |\Psi_0^N\rangle \\
	+ &(a_{I} c_2 - a_{J} c_1) \hat{c}_j |\Psi_0^N\rangle.
\end{split}
\end{equation}
Comparing Eqs.~(\ref{eq:psi0_sudden}) and (\ref{eq:psi0_states}), it is seen that the ionization from the orbital $i$ corresponds to the conditions $a_{I} c_1 + a_{J} c_2=1$ and $a_{I} c_2 - a_{J} c_1=0$ for the expansion coefficients and hole-mixing amplitudes. Solving this system of equations, one finds the expansion coefficients $a_{I}=c_1$ and $a_{J}=c_2$. Similarly, if we ionize an electron from the orbital $j$, the initial state $|\Psi^{N-1}_{j}(0)\rangle$ will be superposition of the ionic states $I$ and $J$ with weights $a_{I}=c_2$ and $a_{J}=-c_1$. Importantly, in the absence of electron correlation, i.e. when each of the ionic states is formed from a single hole configuration only, the hole will stay in the orbital in which it has been initially created and does not migrate. Therefore, many-electron effects play a central role in ultrafast electron dynamics triggered by sudden ionization of a molecule.~\cite{Cederbaum_Zobeley:1999}

\subsection{Charge migration analysis}
\label{sec:charge_migr}
To analyze and visualize the electron motion in a system, an observable property must be computed. Starting from the Born--Huang representation~\cite{Born_Huang:1954} of the molecular wavefunction 
\begin{equation}
	\Theta(\mathbf{r},\mathbf{R},t) = \sum_{I} a_{I}(t) \chi_{I}(\mathbf{R},t) \Psi_{I}(\mathbf{r},\mathbf{R}),
\end{equation}
where $a_{I}(t)$ is, in general, a time-dependent complex amplitude of the electronic state $I$, $\chi_{I}(\mathbf{R},t)$ represents the normalized time-dependent nuclear wave packet propagated on the $I$-th PES, and $\mathbf{r}$ and $\mathbf{R}$ denote electronic and nuclear coordinates, respectively, the expectation value of a general operator $\hat{O}(\mathbf{r},\mathbf{R})$ can be written as
\begin{equation}
\label{eq:expect_value}
\begin{split}
	\langle\hat{O}\rangle(t)= 
		\iint \Theta^{*}(\mathbf{r},\mathbf{R},t)
			\hat{O}(\mathbf{r},\mathbf{R}) \Theta(\mathbf{r},\mathbf{R},t) d\mathbf{r} d\mathbf{R} \\		
		= \sum_{I,J} a^{*}_{I}(t) a_{J}(t) \int
			\chi_{I}^{*}(\mathbf{R},t)O_{IJ}(\mathbf{R}) \chi_{J}(\mathbf{R},t) d\mathbf{R}.
\end{split}
\end{equation}
Here, $O_{IJ}(\mathbf{R})$ are matrix elements of $\hat{O}(\mathbf{r},\mathbf{R})$ operator between electronic states $I$ and $J$
\begin{equation}
	O_{IJ}(\mathbf{R})=
		\int \Psi_{I}^{*}(\mathbf{r},\mathbf{R})
			\hat{O}(\mathbf{r},\mathbf{R})\Psi_{J}(\mathbf{r},\mathbf{R}) d\mathbf{r}.
\end{equation}
If both the operator $\hat{O}(\mathbf{r},\mathbf{R})$ and the electronic states $\Psi_{I}(\mathbf{r},\mathbf{R})$ depend on nuclear coordinates $\mathbf{R}$ only weakly, Eq.~(\ref{eq:expect_value}) for the time-dependent expectation value can be reduced to
\begin{equation}
\label{eq:expect_value_simple}
	\langle \hat{O} \rangle(t) \approx \sum_{I,J} O_{IJ} \chi_{IJ}(t),
\end{equation}
where
\begin{equation}
\label{eq:coherences}
	\chi_{IJ}(t) = a^{*}_{I}(t) a_{J}(t) \int \chi_{I}^{*}(\mathbf{R},t) \chi_{J}(\mathbf{R},t) d\mathbf{R}
\end{equation}
represent the populations of electronic states when $I=J$ and the electronic coherences~\cite{Arnold_Santra:2017,Vacher_Malhado:2017,Despre_Kuleff:2018} when $I \neq J$.

A convenient observable quantity to describe the electron motion triggered by the ionization of a molecule is the hole density, defined as the difference between the stationary electron density of the neutral system before ionization and the time-dependent electron density of the cation~\cite{Cederbaum_Zobeley:1999}:
\begin{equation}
\label{eq:Q_coherences}
\begin{split}
	Q(\mathbf{r},t) & = \langle\Psi_{0}^{N}|\hat{\rho}(\mathbf{r})|\Psi_{0}^{N}\rangle \\
		& -\sum_{I,J} \chi_{IJ}(t)\langle\Psi_{I}^{N-1}|\hat{\rho}(\mathbf{r})|\Psi_{J}^{N-1}\rangle,
\end{split}
\end{equation}
where $\hat{\rho}(\mathbf{r})$ is the one-body electron density operator. Explicit equations for constructing and analyzing the hole density in Eq.~(\ref{eq:Q_coherences}) within the ADC method can be found, e.g., in Refs.~\cite{Breidbach_Cederbaum:2003,Breidbach_Cederbaum:2007}.

\subsection{Thawed Gaussian approximation}
\label{sec:TGA}
As one can see from the general Eq.~(\ref{eq:expect_value_simple}) and a specific example in Eq.~(\ref{eq:Q_coherences}), the time dependence of the expectation value of an electronic operator is determined exclusively by the time dependence of the electronic populations and coherences $\chi_{IJ}(t)$. Evaluation of the quantities $\chi_{IJ}(t)$ via Eq.~(\ref{eq:coherences}) requires, in turn, the knowledge of nuclear wave packets $\chi_{I}(\mathbf{R},t)$ for all the involved states at each moment of time. Here, we perform the wave packet propagation with the TGA,~\cite{Heller:1975} one of the simplest approaches to semiclassical wave packet dynamics in a general potential. We neglect the nonadiabatic transitions in our simulations and thus the populations of electronic states $a_I$ remain constant in time.

The Gaussian wave packet considered in TGA is given in the position representation as
\begin{equation}
\label{eq:TGA_WP}
\begin{split}
	\chi_{I}(\mathbf{R},t)=\exp\bigg\{
		\frac{i}{\hbar} \bigg[
		\frac{1}{2}(\mathbf{R}-\mathbf{R}^{I}_{t})^{T} \cdot \mathbf{A}^{I}_{t} \cdot (\mathbf{R}-\mathbf{R}^{I}_{t}) \\
		+ (\mathbf{P}^{I}_{t})^{T} \cdot (\mathbf{R}-\mathbf{R}^{I}_{t}) + \gamma^{I}_{t}
		\bigg]
	\bigg\},
\end{split}
\end{equation}
where $\mathbf{R}^{I}_{t}$ and $\mathbf{P}^{I}_{t}$ are the phase-space coordinates of the center of the wave packet, $\mathbf{A}^{I}_t$ is a complex symmetric width matrix with a positive-definite imaginary part, and $\gamma^{I}_{t}$ is a complex number whose real part is a dynamical phase and imaginary part ensures the normalization at all times.

In the TGA, the wave packet is propagated in the effective time-dependent potential given by the local harmonic approximation (LHA)
\begin{equation}
\label{eq:LHA}
\begin{split}
	V^{I}_{\text{LHA}}(\mathbf{R},t) = V^{I}(\mathbf{R}^{I}_{t}) 
		+ (\text{grad}_{\mathbf{R}} V^{I} |_{\mathbf{R}^{I}_{t}})^T \cdot (\mathbf{R}-\mathbf{R}^{I}_{t})\\
		+ \frac{1}{2} (\mathbf{R}-\mathbf{R}^{I}_{t})^{T} \cdot 
			\text{Hess}_{\mathbf{R}} V^{I} |_{\mathbf{R}^{I}_{t}} \cdot (\mathbf{R}-\mathbf{R}^{I}_{t})
\end{split}
\end{equation}
of the true potential $V^{I}(\mathbf{R})$ around the center $\mathbf{R}^{I}_{t}$ of the wave packet at time $t$. 

Inserting the wave packet ansatz~(\ref{eq:TGA_WP}) and the effective potential~(\ref{eq:LHA}) into the time-dependent Schr\"odinger equation, we obtain an equivalent system of ordinary differential equations for the evolution of wave packet parameters~\cite{Heller:1975}
\begin{subequations}
\label{eq:TGA_eqs}
\begin{align}
	\dot{\mathbf{R}}^{I}_{t} &= \mathbf{m}^{-1} \cdot \mathbf{R}^{I}_{t}, \label{eq:TGA_R}\\
	\dot{\mathbf{P}}^{I}_{t} &= -\text{grad}_{\mathbf{R}} V^{I} |_{\mathbf{R}^{I}_{t}}, \label{eq:TGA_P}\\
	\dot{\mathbf{A}}^{I}_{t} &= -\mathbf{A}^{I}_{t} \cdot \mathbf{m}^{-1} \cdot \mathbf{A}^{I}_{t} 
		- \text{Hess}_{\mathbf{R}} V^{I} |_{\mathbf{R}^{I}_{t}},\label{eq:TGA_A}\\
	\dot{\gamma}^{I}_{t} &= L^{I}_{t} + \frac{i\hbar}{2} \text{Tr} \left(
		\mathbf{m}^{-1} \cdot \mathbf{A}^{I}_{t}
	\right),\label{eq:TGA_gamma}
\end{align}
\end{subequations}
where $\mathbf{m}$ is the real symmetric mass matrix and $L^{I}_{t}$ denotes the Lagrangian
\begin{equation}
	L^{I}_{t} = \frac{1}{2} (\mathbf{P}^{I}_{t})^{T} \cdot \mathbf{m}^{-1} \cdot \mathbf{P}^{I}_{t} -  V^{I}(\mathbf{R}^{I}_{t}).
\end{equation}
Numerical solution of the system of differential equations~(\ref{eq:TGA_eqs}) is straightforward, although additional transformations can be performed in order to make the integration of Eqs.~(\ref{eq:TGA_A}) and (\ref{eq:TGA_gamma}) more stable (see, e.g., Ref.~\cite{Vanicek_Begusic:2021}). Since the TGA requires only a single classical trajectory, along with the corresponding Hessians, it is particularly suitable for an on-the-fly implementation, where the local properties of the PES are obtained as-needed from an \textit{ab initio} electronic structure calculations. In practice, Eqs.~(\ref{eq:TGA_eqs}) can be solved in three stages: (I) the first two equations [Eqs.~(\ref{eq:TGA_R}) and (\ref{eq:TGA_P})] define a classical trajectory of a particle moving on the corresponding PES, (II) the Hessians are computed in parallel for every position in the classical path and (III) the reconstruction of the width and the phase of the wave packet is performed using Eqs.~(\ref{eq:TGA_A}) and (\ref{eq:TGA_gamma}). 

\subsection{Semiclassical description of decoherence}
\label{sec:SC_analysis}
The simple Gaussian form of the nuclear wave packet~(\ref{eq:TGA_WP}) imposed by the TGA makes it possible to perform the integration step in Eq.~(\ref{eq:coherences}) for the electronic coherences analytically (see, e.g., Ref.~\cite{Begusic_Vanicek2020})
%
%
\begin{equation}
\label{eq:TGA_coherence_full}
\begin{split}
\chi_{IJ}(t) & = a^{*}_{I} a_{J} \sqrt{\frac{(2\pi\hbar)^D}{\text{det}(-i \delta \mathbf{A})}} \\
& \times \exp\bigg\{
	\frac{i}{\hbar} \bigg[
		-\frac{1}{2} \delta \bm{\xi}^{T} \cdot (\delta \mathbf{A})^{-1} \cdot \delta \bm{\xi} 
		+\delta \eta
	\bigg]
\bigg\},
\end{split}
\end{equation}
where we introduced the notation $\delta \Lambda := \Lambda^{J} - (\Lambda^{I})^{*}$ for the difference of tensors $\Lambda^{J}$ and $(\Lambda^{I})^{*}$, $I$ and $J$ denote the corresponding electronic states, and $\Lambda^{I}$ can be the matrix $\mathbf{A}_{t}^{I}$, vector
\begin{equation}
	\bm{\xi}_{t}^{I} = \mathbf{P}_{t}^{I} - \mathbf{A}_{t}^{I} \cdot \mathbf{R}_{t}^{I}
\end{equation}
or scalar
\begin{equation}
	\eta_{t}^{I} = \gamma_{t}^{I} - \frac{1}{2} (\bm{\xi}_{t}^{I} + \mathbf{P}_{t}^{I})^{T} \cdot \mathbf{R}_{t}^{I}.
\end{equation}

Evolution of the width of the wave packet within TGA allows one to account for stretching and compression of the Gaussian along the propagated trajectory. Although this additional flexibility of the wave packet provides, in general, a more accurate solution of the Schr\"odinger equation, it also complicates Eq.~(\ref{eq:TGA_coherence_full}) and thus the analysis of the electronic coherences. Let us assume, for the sake of simplicity, that the widths of the Gaussians propagated in the electronic states $I$ and $J$ remain fixed along trajectories, i.e., $\mathbf{A}^{I}_{t}=\mathbf{A}^{J}_{t}=i\mathbf{\Gamma}$, where $\mathbf{\Gamma}$ is a real positive-definite symmetric matrix. The latter makes it possible to further simplify Eq.~(\ref{eq:TGA_coherence_full}) as (see Ref.~\cite{Golubev_Vanicek:2020})
\begin{equation}
\label{eq:coherence_TGA_simple}
	\chi_{IJ}(t) \approx a^{*}_{I} a_{J} e^{-d(t)^{2}/4\hbar}e^{i \Delta S(t)/\hbar},
\end{equation}
where
\begin{equation}
\label{eq:d}
	d(t)=\sqrt{|\mathbf{\tilde{R}}^{I}_{t}-\mathbf{\tilde{R}}^{J}_{t}|^{2}
		+|\mathbf{\tilde{P}}^{I}_{t}-\mathbf{\tilde{P}}^{J}_{t}|^{2}}
\end{equation}
is the phase-space distance between the centers of the two Gaussian wave packets in mass- and frequency-scaled coordinates and momenta
\begin{subequations}
\begin{align}
	\mathbf{\tilde{R}}^{I}_{t}  &  = \mathbf{\Gamma}^{1/2} \cdot \mathbf{R}^{I}_{t},\\
	\mathbf{\tilde{P}}^{I}_{t}  &  = \mathbf{\Gamma}^{-1/2} \cdot \mathbf{P}^{I}_{t},
\end{align}
\end{subequations}
and the phase term $\Delta S(t)$ is defined as
\begin{equation}
\label{eq:phase_FGA}
	\Delta S(t)= 
		\int_{0}^{t} \left(
			L^{J}_{t'} - L^{I}_{t'}
		\right) dt'
		-\frac{1}{2}
			\left(
				\mathbf{P}^{I}_{t}+\mathbf{P}^{J}_{t}
			\right)^{T} 
		\cdot
			\left(
				\mathbf{R}^{J}_{t}-\mathbf{R}^{I}_{t}
			\right).
\end{equation}
The integral of Laplacian on each surface can be rewritten via Legendre transformation as
\begin{equation}
\begin{split}
	\int_{0}^{t} L^{I}_{t'} dt' = 
		\int_{0}^{t} \left(
		   (\mathbf{P}^{I}_{t'})^T \cdot \frac{d\mathbf{R}^{I}_{t'}}{dt'} - H^{I}_{t'}
		\right) dt' \\
	= \int_{\mathbf{R}_{0}}^{\mathbf{R}^{I}_{t}}
			(\mathbf{P}^{I}_{t'})^T \cdot d \mathbf{R}^{I}_{t'} - t V^{I}(\mathbf{R}_{0}).
\end{split}
\end{equation}
Here, we used the relation $H^{I}(\mathbf{R}^{I}_{t},\mathbf{P}^{I}_{t})=H^{I}(\mathbf{R}_{0},\mathbf{P}_{0})$ because the Hamiltonian is a constant of motion, and $H^{I}(\mathbf{R}_{0},\mathbf{P}_{0}) = V^{I}(\mathbf{R}_{0})$ because $\mathbf{P}_{0}=0$. Then, we can write the phase $\Delta S(t)$ as
\begin{equation}
\label{eq:S_t_res}
	\Delta S(t)=S_{\text{red}}(t)-\Delta Et,
\end{equation}
where
\begin{equation}
\label{eq:delta_E}
	\Delta E=V^{J}(\mathbf{R}_{0})-V^{I}(\mathbf{R}_{0})
\end{equation}
is the energy gap between the electronic states $I$ and $J$ at $\mathbf{R}_{0}$ and
\begin{equation}
\label{eq:S_red}
\begin{split}
	S_{\text{red}}(t) =
		\int_{\mathbf{R}_{0}}^{\mathbf{R}^{J}_{t}}
			(\mathbf{P}^{J}_{t'})^T \cdot d \mathbf{R}^{J}_{t'}
	   -\int_{\mathbf{R}_{0}}^{\mathbf{R}^{I}_{t}}
			(\mathbf{P}^{I}_{t'})^T \cdot d \mathbf{R}^{I}_{t'} \\
	   -\frac{1}{2}
			\left(
				\mathbf{P}^{I}_{t}+\mathbf{P}^{J}_{t}
			\right)^{T} 
		\cdot
			\left(
				\mathbf{R}^{J}_{t}-\mathbf{R}^{I}_{t}
			\right)	
	=\oint_{\mathbf{C}}\mathbf{P}^{T}\cdot d\mathbf{R}
\end{split}
\end{equation}
is the reduced action equal to the signed area within the closed curve $\mathbf{C}$ consisting of the two classical paths connecting $(\mathbf{R}_{0},\mathbf{P}_0)$ with $(\mathbf{R}^{I}_{t},\mathbf{P}^{I}_{t})$ or $(\mathbf{R}^{J}_{t},\mathbf{P}^{J}_{t})$, and a straight line connecting $(\mathbf{R}^{I}_{t},\mathbf{P}^{I}_{t})$ and $(\mathbf{R}^{J}_{t},\mathbf{P}^{J}_{t})$.~\cite{Golubev_Vanicek:2020}

The analytical expression~(\ref{eq:coherence_TGA_simple}) permits a simple semiclassical interpretation of the effect of nuclear dynamics on electronic coherence. The decay of coherence takes place due to the increase of distance between the nuclear wave packets in phase space, Eq.~(\ref{eq:d}). Within the frozen Gaussian approximation the magnitude of the electronic coherence can be interpreted as a product of the wave-packet overlap in coordinate space, i.e., their spatial overlap, and the overlap of the wave packets in momentum space, which is referred to as dephasing.~\cite{Vacher_Malhado:2017} The third mechanism responsible for the decoherence, namely the change of populations of electronic states, is not taken into account in our simulations. Besides the influence on the absolute value of the electronic coherence, the diverging nuclear trajectories affect the frequency of electronic oscillations. In the absence of nuclear motion, the time scale of electronic dynamics is defined by the energy difference between the corresponding electronic states, Eq.~(\ref{eq:delta_E}). Due to the nuclear motion, the static frequency is modified by the area $S_{\text{red}}(t)$ between two nuclear trajectories in the phase space, Eq.~(\ref{eq:S_red}).

Importantly, ignoring the mode--mode mixing (Duschinsky rotation),~\cite{Duschinsky:1937} the term responsible for the decay of the electronic coherence in Eq.~(\ref{eq:coherence_TGA_simple}) can be decomposed into individual coordinate and momentum contributions associated with each of the vibrational modes of the system
\begin{equation}
\label{eq:coh_normal_mode_decomp}
	e^{-d(t)^{2}/4\hbar}=
		\prod_{k}
		e^{
			-(\mathbf{\tilde{R}}^{I}_{t}[k]-\mathbf{\tilde{R}}^{J}_{t}[k])^{2}/4\hbar
		}
		e^{
			-(\mathbf{\tilde{P}}^{I}_{t}[k]-\mathbf{\tilde{P}}^{J}_{t}[k])^{2}/4\hbar
		},
\end{equation}
where the multiplication is made over all the vibrational modes, and $[k]$ denotes $k$-th component of the corresponding vector. As one can see from Eq.~(\ref{eq:coh_normal_mode_decomp}), the loss of wave packets overlap in coordinate or momentum space within a single vibrational mode will lead to the overall decoherence (see also Ref.~\cite{Arnold_Santra:2017}).

\section{Methodology and computational details}
\label{sec:comp_det}

The screening procedure consists of the following four steps: (I) the molecule of interest is downloaded from the PubChem database; (II) the ionic spectrum of the molecule at a fixed reference geometry is computed utilizing the non-Dyson version of the ADC scheme~\cite{Schirmer_Walter:1983,Schirmer_Stelter:1998} at the third order of perturbation theory [nD-ADC(3)]; (III) if the molecule has an appropriate ionic spectrum, namely both sufficient energy gaps between ionic states and strong hole-mixing, an initial prescreening of the electronic coherences is performed using the TGA combined with the TDDFT~\cite{Runge_Gross:1984,Casida:1995}; (IV) the best candidates showing a long lasting coherence are verified by the TGA using the EOM-CC-IP~\cite{Stanton_Bartlett:1993,Kamiya_Hirata:2006} with single and double excitations (EOM-CCSD-IP).

Since the molecules in the PubChem database are collected from different sources, it is hard to estimate the reliability of molecular geometries presented in the database. For this reason, we downloaded molecules in the simplified molecular-input line-entry system (SMILES) format,~\cite{Weininger:1988} which describes only the atoms connectivity in a molecule. Preliminary molecular geometries were constructed using OpenBabel 3.0.0~\cite{Oboyle_Hutchison:2011} with the MMFF94 force field.~\cite{Halgren_Nachbar:1996} Ground-state geometries of the neutral molecules were optimized using the density functional theory (DFT)~\cite{hohenberg1964} at the wB97XD/6-311++G(d,p)~\cite{chai2008} level. The optimization was performed with the Gaussian 16 package.~\cite{Frisch_Fox:2016}

The noncorrelated reference Hartree--Fock (HF) orbitals for subsequent ADC computations were obtained using GAMESS-UK 7.0 package~\cite{Guest_Kendrick:2005} with the standard double-zeta plus polarization (DZP)~\cite{Canal_Jorge:2005} basis set. The active space for every molecule was chosen to contain all the orbitals except $1s$ core states of all heavy atoms. All one-hole ($1h$) and two-hole-one-particle ($2h1p$) electronic configurations were taken into account in the calculations of ionic states and in the follow-up time-dependent density analysis.

The TDDFT prescreening was performed at the wB97XD/6-311++G(d,p) level of theory. The classical nuclear trajectories were propagated with the velocity Verlet algorithm,~\cite{Verlet:1967} where the time step was set at 0.25~fs for a total simulation time of 25~fs. The same classical propagation procedure with PESs computed using the EOM-CCSD-IP method with DZP basis set was applied to verify the TDDFT results for molecules that showed the long lasting electronic coherence. The on-the-fly evaluation of the PESs was done with the Q-Chem package.~\cite{Shao_Head-Gordon:2015}

The time-dependent nuclear wave packets were reconstructed from the classical trajectories using the single-Hessian variant of the TGA method.~\cite{Begusic_Vanicek:2019} The initial wave packets were prepared from the Hessian of the reference ground state of the neutral molecule and distributed between the involved electronic states according to the corresponding hole-mixing weights. The Hessians of the excited states were computed numerically at geometries corresponding to the vertical ionization from the reference ground state.


\section{Results and discussion}
\label{sec:results}
We have applied our search methodology to find polyatomic molecules in which the sudden ionization leads to the appearance of a long lasting electronic coherence. Although various classes of small organic molecules are present in the PubChem database, the hole-mixing is found to appear mostly in molecules containing an electron donor site, such as unsaturated carbon atoms or aromatic rings, and an electron acceptor site, e.g. aldehyde or amine functional groups. The presence of a strong hole-mixing in these types of molecules has been pointed out in previous works on this subject.~\cite{Lunnemann_Cederbaum:2008a,Kuleff_Cederbaum:2010}

The results section is divided into two parts. In Sec.~\ref{sec:mol_from_search}, we first present molecules found in the PubChem database which demonstrate the long lasting electronic coherences as well as clear charge migration oscillations throughout the molecular structure. In Sec.~\ref{sec:mol_from_lit}, we present the simulations of the electronic coherences coupled to nuclear motion in molecules which were previously studied only at a fixed nuclear geometry.

\subsection{Molecules with long lasting electronic coherence}
\label{sec:mol_from_search}

\subsubsection{But-3-ynal}
Let us start with the but-3-ynal molecule which is composed of a chain of four carbon atoms with an alkyne group at one end and an aldehyde group at the opposite end. Interestingly, the oxygen and hydrogen atoms of the aldehyde group are displaced in the opposite directions from the molecular plane formed by the carbon atoms making the molecule asymmetric.

\begin{figure}
\includegraphics[width=\columnwidth]{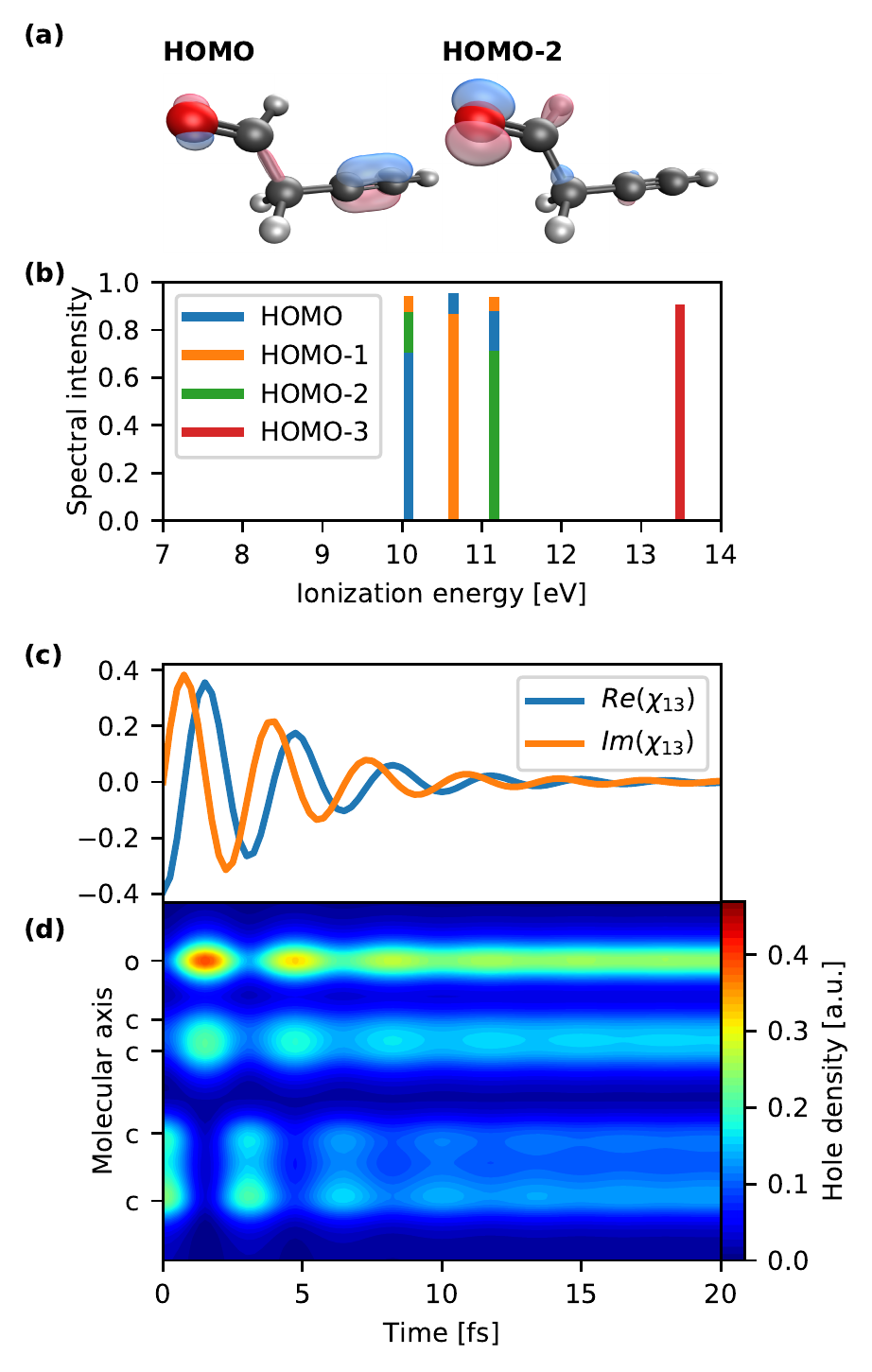}
\caption{The ionization spectrum and the coupled electron-nuclear dynamics triggered by the ionization out of the HOMO of the but-3-ynal molecule. (a) HF molecular orbitals involved in the hole-mixing. (b) First four cationic states computed using nD-ADC(3) method with DZP basis. (c) Time evolution of the electronic coherence between the first and the third cationic states created after removal of the HOMO electron. Dynamics was performed with the semiclassical on-the-fly TGA using EOM-CCSD-IP method. (d) Time evolution of the hole density $Q(z,t)$ along the molecular axis. The charge initially localized in the HOMO orbital migrates back and forth between alkyne and aldehyde moieties of the molecule before being trapped by the nuclear motion.}
\label{fig:butynal}
\end{figure}

The valence molecular orbitals of the neutral but-3-ynal and the ionization spectrum resulting from the removal of an electron from these orbitals are shown in panels (a) and (b) of Fig.~\ref{fig:butynal}, respectively. As one can see, the three lowest ionic states of the molecule are a mixture of one-hole contributions of the valence molecular orbitals. In particular, a strong hole-mixing is seen between the first and the third ionic states: an electron missing in the highest occupied molecular orbital (HOMO) [blue sticks in Fig.~\ref{fig:butynal}(b)] and an electron missing in the HOMO-2 [green sticks in Fig.~\ref{fig:butynal}(b)]. The electron density of the HOMO is localized primarily around the alkyne group, while the HOMO-2 is localized in the vicinity of the aldehyde group. The HOMO-1 [shown by orange sticks in Fig.~\ref{fig:butynal}(b)] is less correlated with the other orbitals and forms an ionic state lying between the pair of states corresponding to the ionization from the HOMO and HOMO-2. A sudden removal of an electron either from HOMO or from HOMO-2 will create an electronic wave packet, which will initiate charge migration oscillations between the carbon triple bond and the aldehyde group with a period of about 3.8~fs, determined by the energy gap between the first and the third cationic states. 

We simulated the ionization from the HOMO of but-3-ynal populating the first and the third electronic states in proportion $\approx$81\%/19\%, respectively, according to the hole-mixing weights. Since the contribution of the HOMO orbital to the second ionic state is marginal, we do not include this state in the initial superposition. The phase between the electronic states is chosen in a way to localize the initial charge in the HOMO. The evolution of electronic coherence created by such a superposition is shown in Fig.~\ref{fig:butynal}(c). As one can see, the oscillations of the coherence gradually dephase within 10~fs due to the coupling to the nuclear motion. The period of the oscillations is slightly different from that predicted with the static nuclei. This is partially due to the difference between ADC and EOM-CC-IP electronic energies (see Sec.~\ref{sec:comp_det} for details) but also reflects the influence of nuclear motion on the time scale of electronic oscillations [see Eqs.~(\ref{eq:S_t_res}) and (\ref{eq:S_red}), and also Ref.~\cite{Golubev_Vanicek:2020}].

The hole density $Q(z,t)$ computed along the molecular axis of the but-3-ynal is shown in Fig.~\ref{fig:butynal}(d). The quantity $Q(z,t)$ was obtained by integrating $Q(\mathbf{r},t)$, Eq.~(\ref{eq:Q_coherences}), over the $x$ and $y$ components, perpendicular to $z$, while the axis $z$ was chosen to pass through the longest spatial extension of the molecule, thus called ``molecular axis''. As one can see from Fig.~\ref{fig:butynal}(d), the charge performs a couple of clear oscillations before being trapped by the nuclear motion and distributed along the molecular chain.

\begin{figure}
\includegraphics[width=\columnwidth]{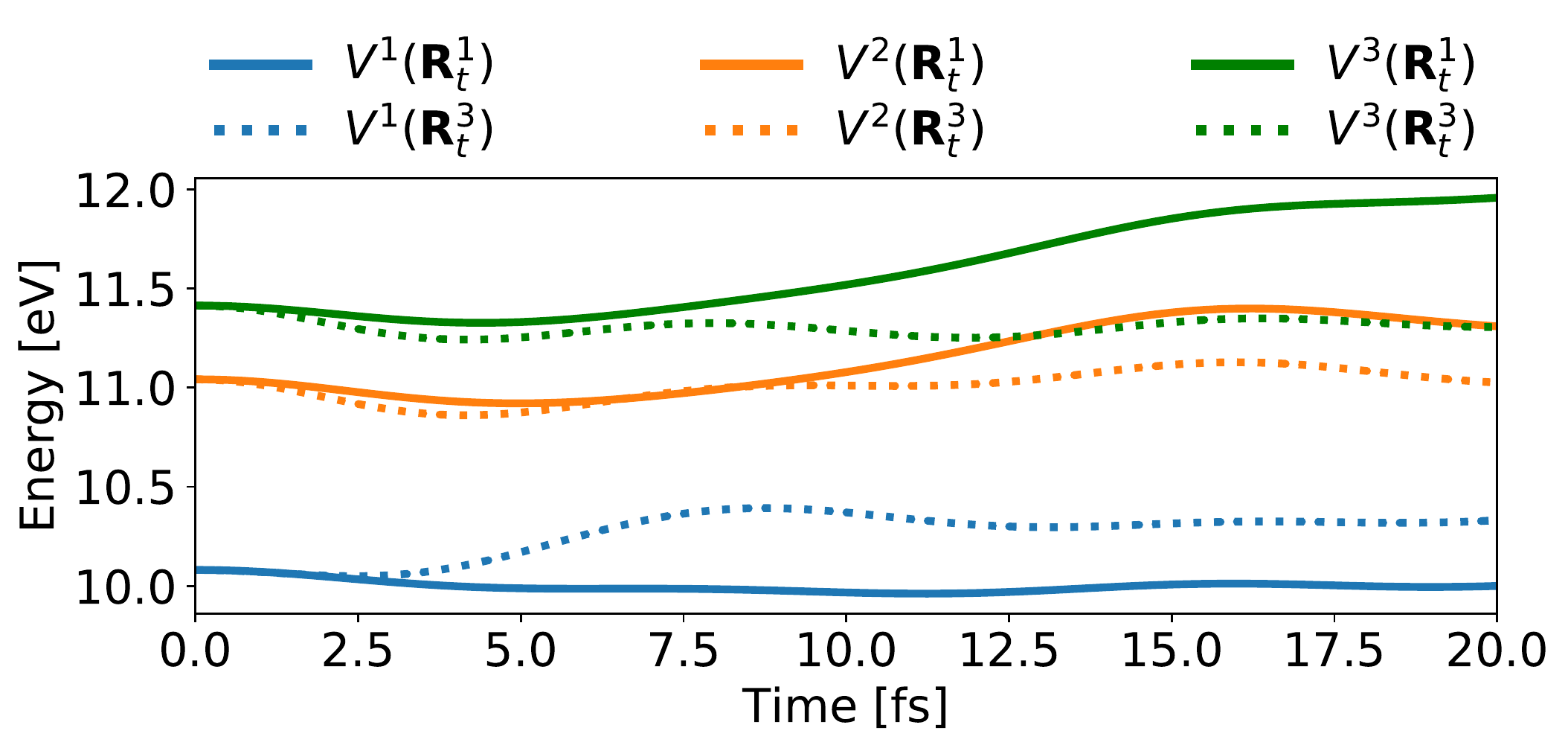}
\caption{Energies $V^{I}(\mathbf{R}_{t}^{J})$ of the $I$-th cationic state along the nuclear trajectory propagated on the  $J$-th PES of the but-3-ynal molecule. The sufficient energy gaps between the involved states along both trajectories justify the negligible influence of the nonadiabatic effects on the dynamics.}
\label{fig:butynal_PES}
\end{figure}

To ensure that the nonadiabatic effects do not lead to a significant population transfer between the neighboring electronic states, we monitored the energies of the three lowest states along the trajectories propagated in the first and the third states. As one can see from Fig.~\ref{fig:butynal_PES}, the energy gaps between the states along each of the trajectories are sufficiently large to neglect the nonadiabatic transitions. Note that, in general, the population transfer can occur also between energetically distant adiabatic electronic states due to the momentum term in the Hamiltonian (see, e.g., Ref.~\cite{Choi_Vanicek:2021} for more details). However, in our scheme the initial wave packets have zero momentum at the beginning of the propagation and the propagation time is short enough to assume the negligible influence of the nonadiabatic effects caused by the momentum operator.

%

\subsubsection{Pent-4-enal}
\label{subsec:pentenal}
The pent-4-enal molecule is structurally similar to but-3-ynal but is composed of a chain of five carbon atoms with an alkene group at one end and an aldehyde group at the opposite end. Due to the lack of symmetry the molecule belongs to the C$_{1}$ point group.

\begin{figure}
\includegraphics[width=\columnwidth]{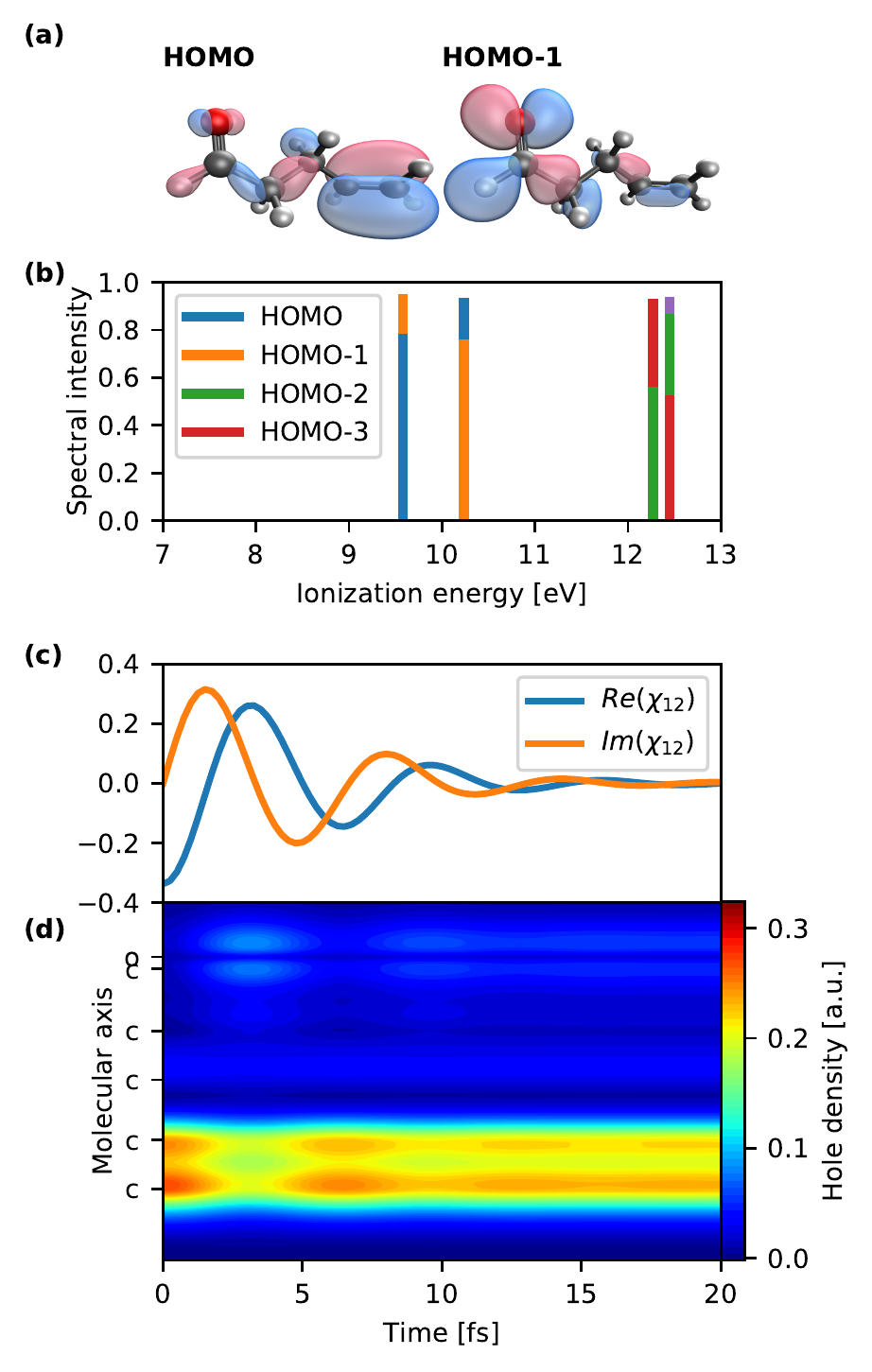}
\caption{The ionization spectrum and the coupled electron-nuclear dynamics triggered by the ionization out of the HOMO of the pent-4-enal molecule. See the caption of Fig.~\ref{fig:butynal} for the explanation of the four panels.}
\label{fig:pentenal}
\end{figure}

The valence molecular orbitals of the neutral pent-4-enal and the ionization spectrum are shown in panels (a) and (b) of Fig.~\ref{fig:pentenal}, respectively. Similarly to the but-3-ynal, the strong correlation between valence electrons of the molecule leads to the appearance of the hole-mixing between the ionic states. As one can see, the two lowest ionic states are a mixture of the one-hole contributions of the HOMO and HOMO-1, while the third and the fourth states are composed from the HOMO-2 and HOMO-3.

We simulated the ionization from the HOMO populating the first and the second electronic states in proportion $\approx$87\%/13\%, respectively, according to the hole-mixing weights. The evolution of electronic coherence created by such a superposition is shown in Fig.~\ref{fig:pentenal}(c). Similarly to the but-3-ynal, the electronic coherence dephases within 10~fs due to the nuclear rearrangement. However, since the energy gap between the involved states in the case of pent-4-enal is smaller than that of the but-3-ynal, the electronic coherence performs a smaller number of oscillations within the same time. As for the ionization from the HOMO-2 and HOMO-3, the small energy gap between the third and the fourth ionic states makes the oscillations of the electron density so slow that the decoherence takes place faster than the period of a single oscillation.

The hole density $Q(z,t)$ computed along the molecular axis of the pent-4-enal is shown in Fig.~\ref{fig:pentenal}(d). As one can see, the charge oscillates between the double bond and the aldehyde group of the molecule before being trapped by the nuclear motion.

%

\subsubsection{2,5-dihydrofuran and 3-pyrroline}
Interesting examples of a strong electron correlation between an unsaturated carbon bonds and an electron acceptor atom can be found in the cyclic molecules 2,5-dihydrofuran and 3-pyrroline. Both structures contain a five-atom ring with a heteroatom (oxygen in 2,5-dihydrofuran and nitrogen in 3-pyrroline) opposite to a carbon double bond. The 2,5-dihydrofuran molecule belongs to the C$_{\text{2v}}$ symmetry, while the additional hydrogen atom in 3-pyrroline reduces its symmetry to the C$_{\text{s}}$ point group.

\begin{figure}
\includegraphics[width=\columnwidth]{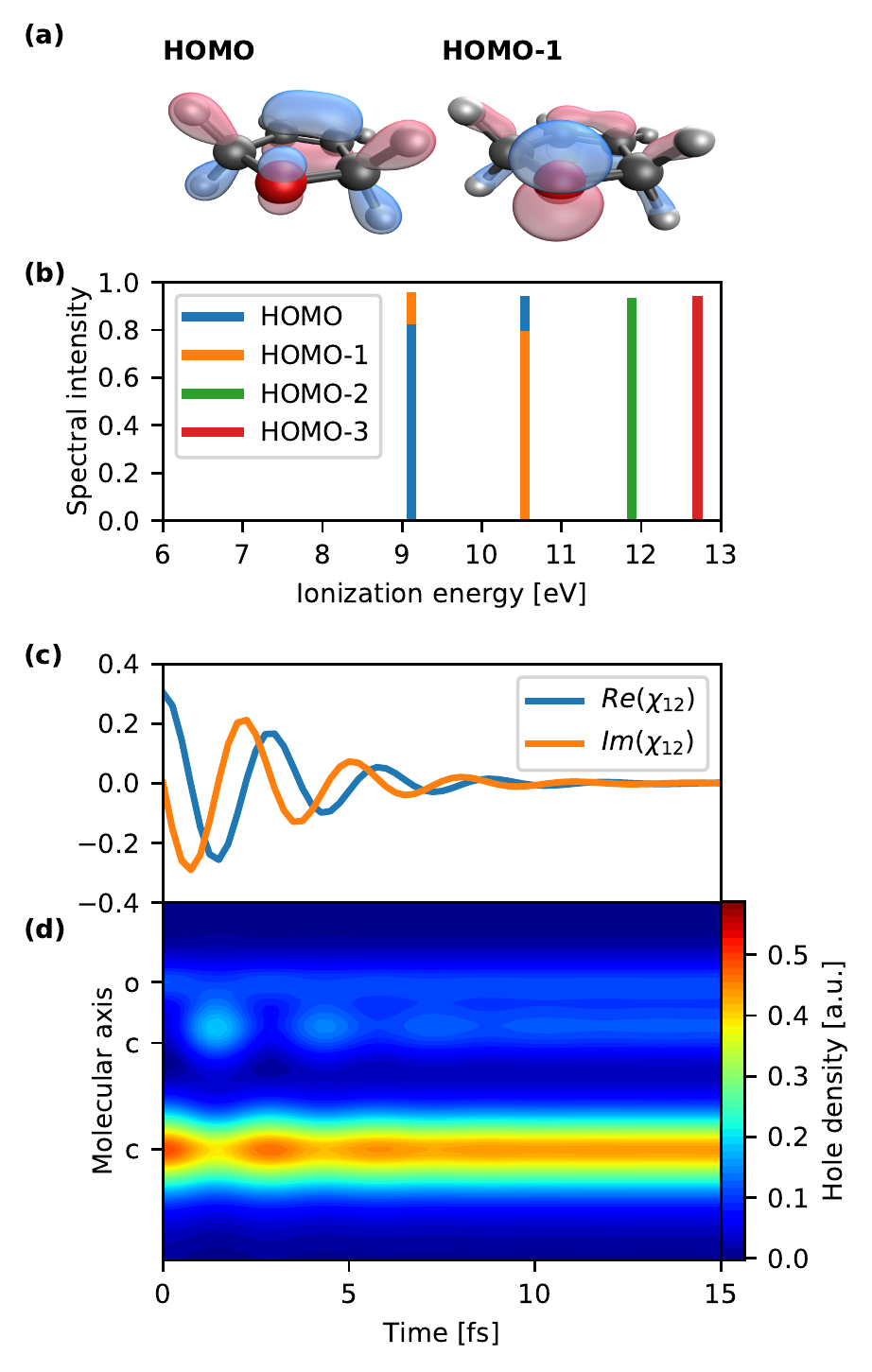}
\caption{The ionization spectrum and the coupled electron-nuclear dynamics triggered by the ionization out of the HOMO of the 2,5-dihydrofuran molecule. See the caption of Fig.~\ref{fig:butynal} for the explanation of the four panels.}
\label{fig:dihydrofuran}
\end{figure}

\begin{figure}
\includegraphics[width=\columnwidth]{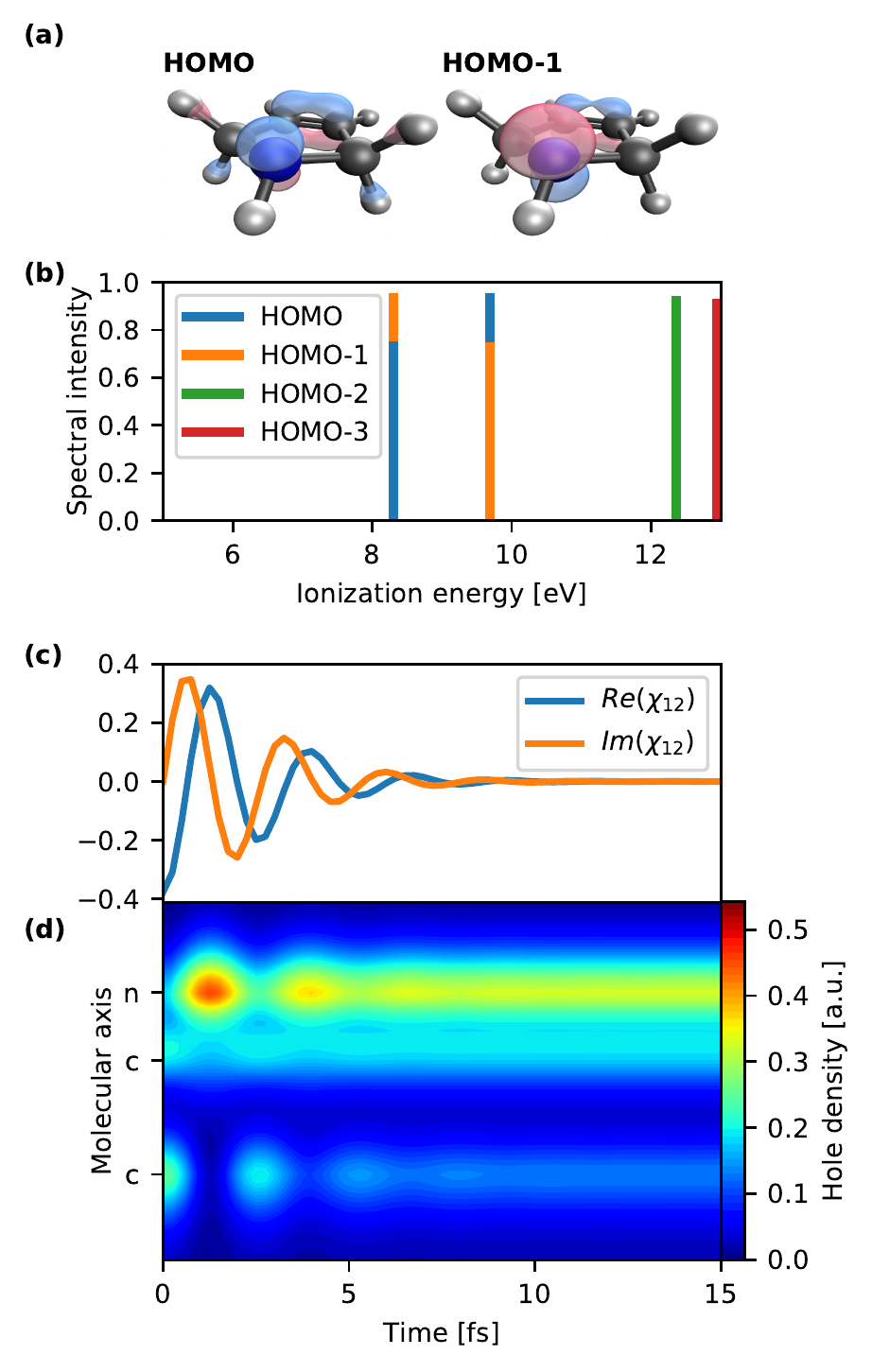}
\caption{The ionization spectrum and the coupled electron-nuclear dynamics triggered by the ionization out of the HOMO of the 3-pyrroline molecule. See the caption of Fig.~\ref{fig:butynal} for the explanation of the four panels.}
\label{fig:pyrroline}
\end{figure}

The valence molecular orbitals of the neutral molecules and the ionization spectra of 2,5-dihydrofuran (Fig.~\ref{fig:dihydrofuran}) and 3-pyrroline (Fig.~\ref{fig:pyrroline}) are shown in panels (a) and (b), respectively. As one can see, in both cases the first two cationic states consist of a strong mixture of the HOMO and HOMO-1 in similar proportions. The HOMOs of both molecules are localized in the double bond while the HOMOs-1 are localized in the vicinity of the heteroatom.

The electronic coherences appearing after the removal of the HOMO electron from 2,5-dihydrofuran and 3-pyrroline are shown in panels (c) of Figs.~\ref{fig:dihydrofuran} and \ref{fig:pyrroline}, respectively. In both molecules the coherence performs several oscillations before being destroyed by the nuclear rearrangement. The hole densities $Q(z,t)$ computed along the molecular axes passing through the midpoint of the double bond and the heteroatom are shown in panels (d) of Figs.~\ref{fig:dihydrofuran} and \ref{fig:pyrroline}. In both molecules the charge initially localized in the double bond oscillates from one side of the system to the other with a period of about 3~fs before being trapped and distributed along the molecule.

The 2,5-dihydrofuran and 3-pyrroline are interesting candidates for experimental studies involving X-ray transient absorption measurements~\cite{Golubev_Kuleff:2021} because of the different heteroatoms contained in the molecules. By taking advantage of element-specific core-to-valence transitions induced by X-ray radiation, one can trace the dynamics of electron density with atomic spatial resolution. Depending on the available laser setup, the energy window corresponding to the absorption by the O or N atom in 2,5-dihydrofuran and 3-pyrroline, respectively, can be exploited.

%

\subsection{Molecules from the literature}
\label{sec:mol_from_lit}
Let us turn to the analysis of the electron-nuclear dynamics in 2-propyn-1-ol, PENNA, BUNNA and MePeNNA molecules which were previously studied employing the frozen nuclei approximation (see Refs.~\cite{Breidbach_Cederbaum:2007,Lunnemann_Cederbaum:2008,Lunnemann_Cederbaum:2008a}). Similarly to most of the molecules found in the present study, the strong electron correlation between molecular orbitals localized around donor and acceptor sites of the neutral 2-propyn-1-ol, PENNA, BUNNA and MePeNNA leads to the appearance of the hole-mixing in the ionic states. The ionization spectra of all the molecules are somewhat similar~\cite{Breidbach_Cederbaum:2007,Lunnemann_Cederbaum:2008,Lunnemann_Cederbaum:2008a} and represent a few lines corresponding to the ionization of the outer-valence electrons separated by an energy gap from the remaining ionic states. To simplify the comparison of the electronic coherence for different molecules, here we assume that only two states with largest hole-mixing amplitudes become equally populated for each of the studied systems.

\begin{figure}
\includegraphics[width=\columnwidth]{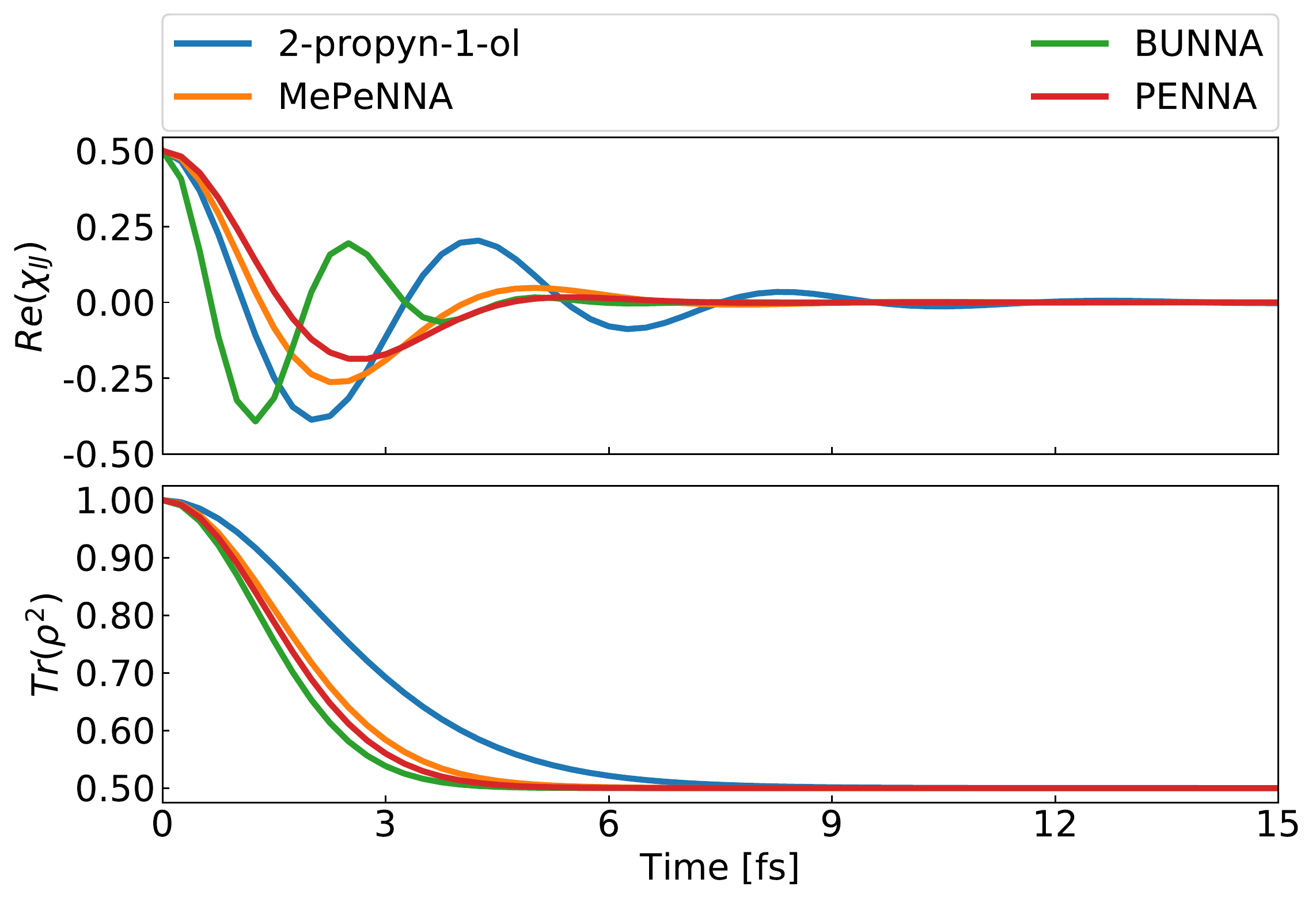}
\caption{Top panel: Time evolution of the real part of electronic coherences created between $I$ and $J$ ionic states of 2-propyn-1-ol ($I=1$, $J=3$), BUNNA ($I=1$, $J=2$), PENNA ($I=1$, $J=2$), and MePeNNA ($I=1$, $J=2$). Bottom panel: Comparison of the electronic purity functions $\text{Tr}[\rho(t)^2]$ computed for the specified combinations of the electronic states.}
\label{fig:propynol_MePeNNA_BUNNA_coherence}
\end{figure}

Figure~\ref{fig:propynol_MePeNNA_BUNNA_coherence} shows the electronic coherence in the four considered molecules for various combinations of the electronic states. A convenient quantity allowing a comparison of the coherence times of different molecules is the purity function $\text{Tr}[\rho(t)^2]$, where the electron density matrix $\rho(t)$ is related to the matrix of nuclear overlaps, Eq.~(\ref{eq:coherences}), by transposition: $\rho_{IJ}(t) = \chi_{JI}(t)$. Due to decoherence, the purity decays from the value $\text{Tr}[\rho(0)^2]=1$ for the initially pure state to the value $1/n$ for the equally weighted mixture of $n$ states. The bottom panel of Fig.~\ref{fig:propynol_MePeNNA_BUNNA_coherence} demonstrates that the electronic coherences in BUNNA, PENNA and MePeNNA are fully suppressed within about 5~fs (green, red and orange solid lines in Fig.~\ref{fig:propynol_MePeNNA_BUNNA_coherence}, respectively). Due to the small energy gaps between the involved electronic states, the electron density has time to perform only a single oscillation in the case of PENNA and MePeNNA molecules (see the top panel of Fig.~\ref{fig:propynol_MePeNNA_BUNNA_coherence}). Although the decoherence takes place on the similar time scale in the BUNNA molecule, the large energy gap between the states makes it possible to observe more oscillations of the electron density in this case. The electronic coherence in the 2-propyn-1-ol survives the nuclear motion for about 8~fs which makes it possible to observe a couple of oscillations of the electron density in this molecule.

\section{Conclusions and outlook}
\label{sec:conclusions}
In this paper, we have performed a thorough theoretical scan of a large number of small polyatomic molecules searching for specific structural and dynamical properties which can be useful for the experimental measurements of ultrafast electronic dynamics and its coupling to the nuclear motion. In particular, we were interested to find molecules with long-lasting electronic oscillations, surviving the decoherence caused by the nuclear rearrangement for the longest possible time. We concentrated on studying the correlated many-electron dynamics induced by the outer-valence sudden ionization with short intense laser pulses. Using a combination of high-level electronic structure techniques with efficient on-the-fly semiclassical description of nuclear dynamics, we have analyzed in total about 250 molecules with potentially interesting properties. Our results show that in most of the studied molecules, the sudden ionization either does not lead to a superposition of states trough the hole-mixing mechanism, or the obtained electronic coherences become damped by the slow nuclear dynamics on a time scale of a few femtoseconds, which can make experimental measurements of laser-induced electron motion in these systems problematic. Yet, we have found a long lasting electronic coherences in several new molecules, which have thus become promising candidates for experimental studies. In addition, we have performed the full-dimensional simulations of electron-nuclear dynamics in 2-propyn-1-ol, BUNNA, PENNA, and MePeNNA molecules which were previously studied only in the static nuclei approximation. We observed that the electronic coherence in the 2-propyn-1-ol lasts for about 8~fs before being captured by the nuclear motion, while in PENNA, BUNNA and MePeNNA the electron dynamics is fully suppressed within 5~fs.

We would like to emphasize again that we have concentrated here on studying only those molecules which demonstrated specific features in their ionic spectra. In particular, the dynamical simulations were carried out only for the molecules with a strong hole-mixing between the lowest ionic states. At the same time, there exist various other mechanisms of the correlation-driven ultrafast electron dynamics in molecules~\cite{Breidbach_Cederbaum:2003,Kuleff_Cederbaum:2014} which we do not investigate in the present study. In addition, we excluded molecules with the strong nonadiabatic effects which can not be accurately treated within the employed computational scheme. These systems, however, can be promising candidates for studying the appearance of electronic coherences in the vicinity of a conical intersection,~\cite{Kowalewski_Mukamel:2015} as well as for exploring the transfer of electronic coherence between electronic states due to the nonadiabatic processes.~\cite{Matselyukh_Worner:2021} Nonetheless, all of the reported molecules demonstrating a long lasting electronic coherence are readily available in the market and, thus, can be useful candidates for studying the ultrafast many-electron charge migration dynamics.

The semiclassical vertical-Hessians TGA used in this paper can be further improved by calculating Hessians along the propagated trajectory and thus take into account more complicated situations, e.g., dissociation of a molecule. Moreover, the improved versions of the TGA such as the extended thawed Gaussian approximation (ETGA),~\cite{Patoz_Vanicek:2018} which propagates a Gaussian wave packet multiplied by a general polynomial, or a so-called three thawed Gaussians approximation (3TGA),~\cite{Begusic_Vanicek:2018} benefiting from representation of the wave packet by multiple Gaussians, were recently reported which can make on-the-fly semiclassical simulations even more accurate.

Finally, we would like to point out that the efficient approach used in this work opens the door to the analysis of electron-nuclear dynamical processes in larger, biologically relevant systems. Being able to treat molecules with a few hundred atoms, the TGA technique combined with the appropriate electronic structure method can help shed light on the continuing debates on the role of quantum coherence in biology,~\cite{engel2007,collini2010,panitchayangkoon2010} quickly preselect molecules suitable for further experimental investigations, and support theoretically recent experimental observations of attosecond electron dynamics in realistic molecular systems. We hope that our work will motivate such studies.

\section*{Acknowledgments}
The authors wish to thank Alexander Kuleff for numerous valuable discussions and acknowledge financial support from the Swiss National Science Foundation through the National Center of Competence in Research MUST (Molecular Ultrafast Science and Technology) and from the European Research Council (ERC) under the European Union's Horizon 2020 research and innovation programme (Grant Agreement No. 683069--MOLEQULE). N.\ V.\ G. acknowledges the support by the Branco Weiss Fellowship---Society in Science, administered by the ETH Z\"urich.

The authors have no conflicts to disclose.

\section*{Data availability}
The data that support the findings of this study are available from the corresponding author upon reasonable request.

\section*{References}
\bibliographystyle{apsrev4-1}
\bibliography{biblio52,additions_SearchLongCoherence}

\end{document}